\documentclass[twocolumn,letterpaper,aps,prc,longbibliography,superscriptaddress,showpacs,nofootinbib,floatfix]{revtex4-1}

\usepackage{graphicx}
\usepackage{xspace}
\usepackage{amsmath}
\usepackage{color}
\usepackage{hyperref}

\usepackage{lineno}

\newcommand{\Npart} {\ensuremath{N_{\rm part}}}
\newcommand{\snn}          {\ensuremath{\sqrt{s_{\mathrm{NN}}}}}
\newcommand{\sqrtsnn}[1]   {\ensuremath{\sqrt{s_{\mathrm{NN}}}=#1~\mathrm{TeV}}}

\newcommand{\sqts}         {\ensuremath{\sqrt{s}}}
\newcommand{\Ethird}       {\ensuremath{\snn^{1/3}}}
\newcommand{\dNdetaNparthalf} {\ensuremath{{\rm d}N_{\rm ch} / {\rm d}\eta / N_{\rm part} / 2}}

\begin{document}
\title{On the upper bound of entropy production rate from particle multiplicity in heavy ion collisions}

\newcommand{\lbl}{Lawrence Berkeley National Laboratory, Berkeley, California 94720, USA}
\newcommand{\bratislava}{Institute of Physics, Slovak Academy of Sciences, Dubravska cesta 9, 84511 Bratislava, Slovakia}
\affiliation{\lbl}
\affiliation{\bratislava}
\author{M.~Ploskon} \affiliation{\lbl}
\author{M.~Veselsky} \affiliation{\bratislava}

\begin{abstract}
We provide a simple derivation for particle production in heavy-ion collisions that is proportional to the rate of entropy production. We find that the particle production depends only on the power of the centre-of-mass collision energy \snn\ and the effective phase-space/volume (e.g. geometry of the collision approximated by the number of nucleons participating in the collision \Npart). We show that at low-energies the pseudo-rapidity density of particles per participating nucleon pair scales linearly with \snn\ while at high-energies with \Ethird. The \Ethird\ region is directly related to sub-nucleon degrees of freedom and creation of a quark-gluon plasma (QGP). This picture explains experimental observation that the shape of the distributions of pseudorapidity-density per nucleon pair of charged particles does not depend on \snn\ over a large span of collision energies. We provide an explanation of the scaling and connect it with the maximum rate per unit time of entropy production. We conclude with
 remarks on the hadron-parton phase transition. In particular, our considerations suggest that the pseudo-rapitidy density of the produced particles per \Npart/2 larger than approximately 1 (excluding particles from jet fragmentation) is a signature of a QGP formation.
\end{abstract}

\maketitle

\section{Rate of entropy production in heavy-ion collisions}

Charged particle multiplicity data from nucleus-nucleus collisions obtained in last decades at SPS, RHIC and LHC show an interesting feature.
The dependence of charged particle multiplicity on number of participating nucleons \Npart\ does not change from \sqrtsnn{0.008} \cite{Adare:2015bua} to \sqrtsnn{5.02} \cite{Adam:2015ptt}.
Such trend suggests that despite the large differences in centre-of-mass energy the production of particles is govern by similar underlying production mechanisms and the phase space available for particle production characterized by \Npart.
However, while the \dNdetaNparthalf\ distributions are similar the total multiplicity grows with collision energy as the cross-section for inelastic processess increases.
Indeed, the increase of the multiplicity can be seen as an increase of the energy dissipated into the created system.
In inelastic nucleon-nucleon (parton-parton) collisions the total dissipated energy can be expected to grow linearly with center-of-mass energy (e.g. since average energy lost in elastic two body collision is $\sqts/2$).
Consequently, the total energy of the formed expanding fireball (eventually at high-energies consisting of a deconfined matter) for a given \Npart, and due to \Npart\ dependence identical for all beam energies, is expected to be proportional to the centre-of-mass energy.
This dissipated energy drives the expansion of the volume of the fireball until the energy density reaches a critival value at which point so-called freeze-out occurs, e.g. all partons are confined into hadrons.
Since the energy dissipated into the fireball is proportional to beam energy, the frezeout volume is also proportional to beam energy and its radius will be proportional to \Ethird.
Furthermore it can be assumed that the process of hadronization, as described e.g. by Lund model \cite{1979ZPhyC...1..105A,LundModel}, proceeds by breaking the strings along the beam direction once they reach a fixed distance, close to the range of strong force, approximately 1 fm.
Thus starting from a given number of partons (quarks and gluons), each of them will produce a number of hadrons proportional to the distance traveled during one-dimensional expansion of the fireball.
Therefore, at sufficiently high collision energies the number of produced hadrons will be proportional to one-dimensional phase space which, by comparison to two-body collision scenario, can be estimated as proportional to \Ethird.
In the context of entropy production, such scenario implies a fixed rate of particle (and thus entropy) production per time.

\section{Transition from three-dimensional to one-dimensional phase space as a signal of phase transition.}

At lower energies below the onset of deconfinement, production of charged particles occurs via nucleon-nucleon collisions. E.g. when colliding two nuclei (in center of mass frame), the energy available for production of emitted particles in nucleon-nucleon collisions is proportional to beam energy in the c.m. frame $\snn$. Thus it appears natural that their number will scale with beam energy. Of course also available phase space will play role, due to essentially isotropic distribution of collision products it will be a three-dimensional phase space (in fact a product of three-dimensional volume and three-dimensional momentum space). So in the regime, where production of secondary particle (for simplicity we assume that they are dominantly pions) is not hindered by limitations of phase space due to emission threshold, one can expect that

\begin{equation}
N_{\pi {\rm 3D}} = f \frac{\snn}{\left<E_{\pi}\right>} = g V_{0} \Omega_{\rm 3D} \frac{1}{h^{3}}
\end{equation}

where $f$ is a factor reflecting relative energy loss in nucleon-nucleon collisions and $g$ is a degeneration factor, $V_{0}$ is volume where pions are produced, $\Omega_{\rm 3D}$ is volume of available momentum space, and $\left<E_{\pi}\right>$ is the mean energy of produced particles.
Of these, $V_{0}$ can be estimated from two-pion Bose-Einstein correlations (HBT) \cite{Aamodt:2011mr} and it can be simplified as a cube with a side length $L$ (in the range of few to about 12 fm depending on the collision energy).
$\Omega_{\rm 3D}$ can be also estimated from experimental momentum distribution (and it is related also to $\left<E_{\pi}\right>$).
This regime with $N_{\pi}$ proportional to $\snn$ can be expected for hadronic gas.

\begin{figure}[tb]
\begin{center}
\includegraphics[width=8cm]{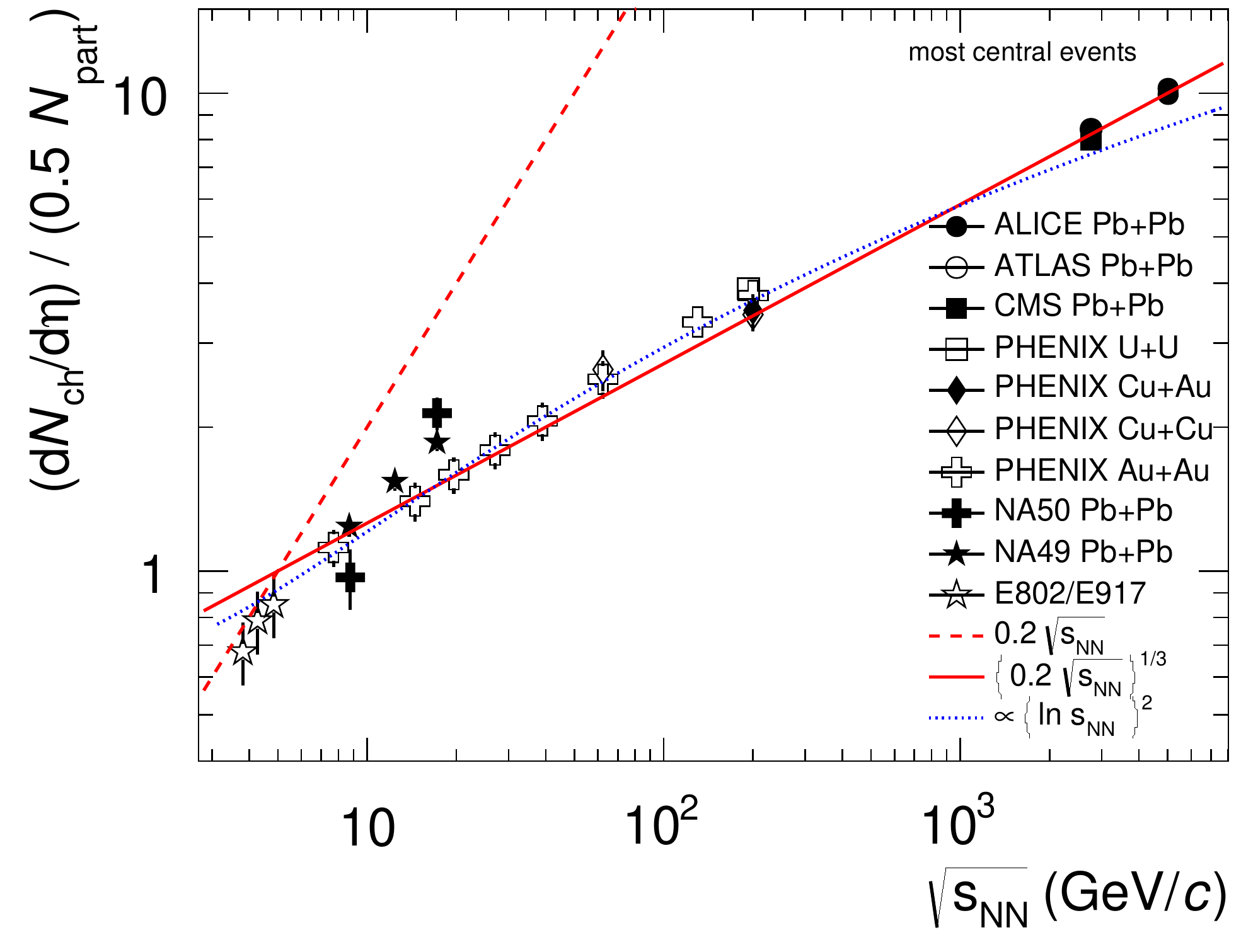}
\caption{Excitation function of pseudorapidity density of charged particles per participating nucleon pair in heavy-ion collisions \cite{Afanasiev:2002fk,Ahle:1999jm,Ahle:2000wq,Abreu:2002fw,Aamodt:2010cz,Adam:2015ptt,ATLAS:2011ag,Chatrchyan:2011pb,Adler:2004zn,Adare:2015bua}. Data points are for most central collisions. See text for more details.}
\label{logmult}
\end{center}
\end{figure}

\begin{figure}[tb]
\begin{center}
\includegraphics[width=8cm]{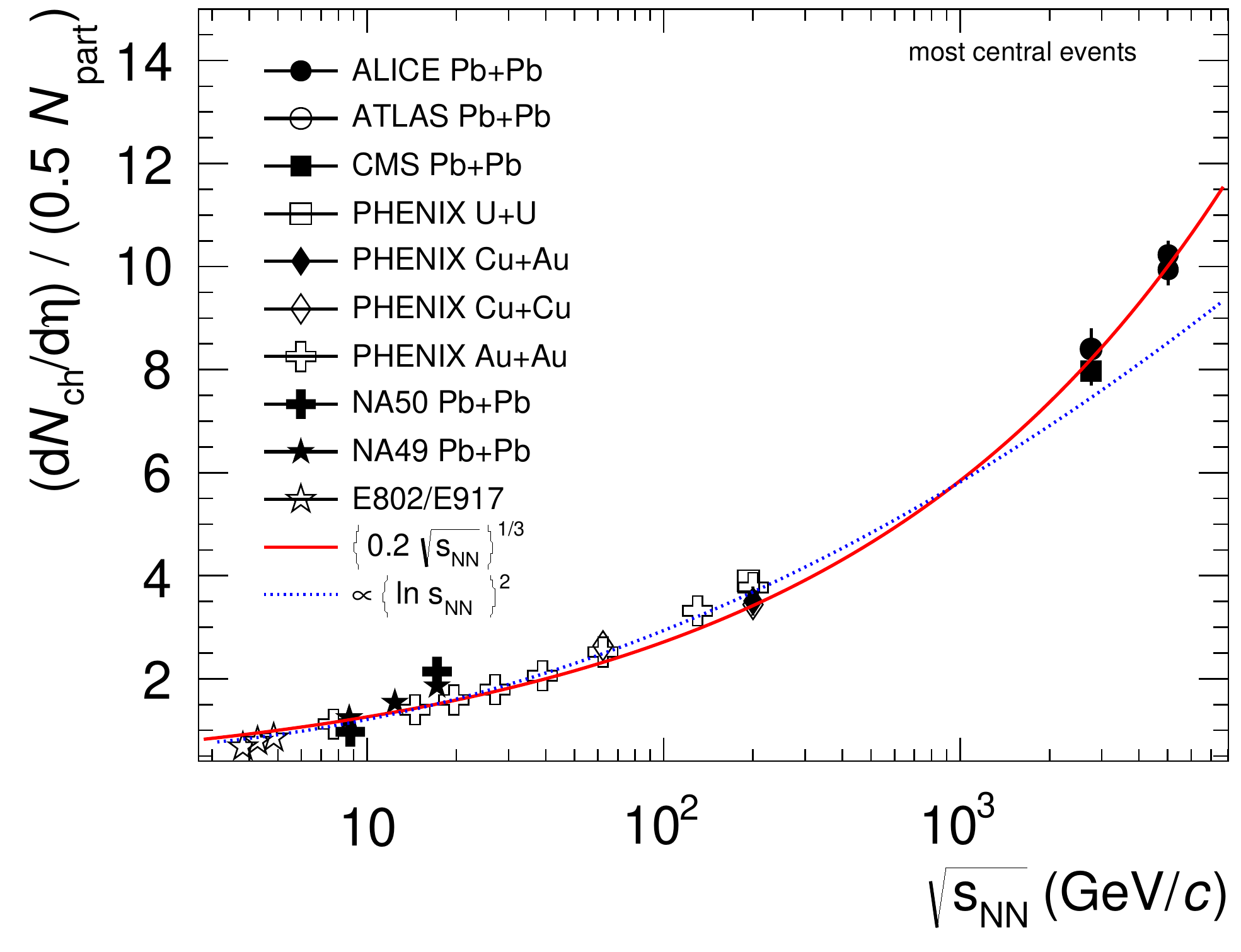}
\caption{Same as Fig. \ref{logmult} but shown with linear multiplicity axis emphasizing the comparison of the parametrizations of the excitation functions at the LHC energies.}
\label{linearmult}
\end{center}
\end{figure}

Indeed, as shown in Fig. \ref{logmult} the experimental data on ${\rm d}N_{\rm ch}/{\rm d}\eta$ tend to follow $\snn$-scaling in the region below 7 GeV, for the AGS (E802/E917) \cite{Ahle:1999jm,Ahle:2000wq} data up to the SPS \cite{Afanasiev:2002fk,Abreu:2002fw} and low-energy RHIC \cite{Adare:2015bua}.
We acknowledge the results from SIS by FOPI Collaboration \cite{Reisdorf:1996qj} but we do not consider those as they are too close to pion emission threshold.

The observed $\snn$  scaling validates the 3D phase space scenario described above.
However, at higher energies, at about 10 GeV the $\snn$-scaling clearly breaks down and experimental multiplicity starts to grow much slower.
It appears, that scenario considering only two-body collisions is no longer valid there.
As an alternative to two-body collision scenario it can be assumed, that at high energies pions are produced by fragmentation of strings between two participant quarks, in the fashion described e.g. by Lund model \cite{1979ZPhyC...1..105A,LundModel}.
This implies that while available phase space does not change dramatically, it will be filled in a different way.
The occupation of space in transverse directions is determined by the number of participating quarks, and only the direction parallel to the beam represents the available phase space.
Then one can expect that number of pions will be expressed as

\begin{equation}
N_{\pi{\rm 1D}} = \frac{1}{h} \frac{N_{f{\rm part}}}{2} g L \Omega_{\rm 1D}
\end{equation}

where for each of $N_{f{\rm part}}/2$ strings one-dimensional phase space $(g L \Omega_{\rm 1D}) / h$ is filled. This is a scenario which can be expected in nucleus-nucleus collisions where quark-gluon plasma is created.
To make a meaningful comparison of these two reaction and phase-space scenarios, one can utilize the fact that such phase-space is constrained experimentally by HBT volume and momentum distribution, and this constraint can be applied to both scenarios in order to estimate $N_{\pi}$.
Since volume can be approximated as $V_{0} = L^{3}$ and volume of momentum space as $\Omega_{\rm 3D} = \Omega_{\rm 1D}^{3}$ one has the following relation between three-dimensional and one-dimensional phase space

\begin{equation}
V_{0} \Omega_{\rm 3D} = (L \Omega_{\rm 1D})^{3}
\end{equation}

where three-dimensional phase space is a cube of one-dimensional phase-space.
As discussed above, at energies below 10 GeV the 3D phase space scales with beam energy.
This also implies that each of the three components scales with \Ethird.
It is straight forward to assume that this scaling is preserved also above 10 GeV, such that

\begin{equation}
N_{\pi {\rm 1D}} = \frac{N_{f {\rm part}}}{2} g \left(\frac{V_{0} \Omega_{\rm 3D}}{h^{3}}\right)^{1/3} =  \frac{N_{f {\rm part}}}{2} g \left(f \frac{\snn}{\left<E_{\pi}\right>}\right)^{1/3}
\end{equation}

what implies, even if only indirectly, by comparison to hadronic gas scenario, that in the QGP regime the charged particle multiplicity ought to scale with \Ethird.

The value of $\left<E_{\pi}\right>$ in 3D phase space scenario differs from the same quantity in 1D scenario only by a scaling factor.
The above trend is indeed observed experimentally, as shown in Fig. \ref{logmult} for most central collisions.
The  $\sqts^{1/3}$ scaling (solid line) is observed in the region ranging from the SPS, through RHIC, up to the data from the LHC.
Moreover, as noted earlier, in the whole region of \snn\ between 10 GeV and 5.02 TeV the shapes of \Npart-dependence of multiplicity are essentially identical and this along with $N_{f {\rm part}}$ scaling supports the 1D geometry of such phase space scenario.
The agreement with the simple \Ethird\ scaling over the broad range of energies leads us to claim that the formation of QGP is present in collisions from \sqrtsnn{0.01} on.
Moreover, the transition between the two, \snn\ and \Ethird\ regimes is correlated with the deconfinement phase transition.

\section{Discussion}

\subsection{On figures and excitation functions}

In Fig. \ref{logmult} and Fig. \ref{linearmult} most of the data are for either 5\% or 10\% of the cross-section except ATLAS that is for 6\% most central collisions.
These are compared to curves parametrizing different dependency on centre-of-mass energy.
Two domains of particle production are observed: one at energies below about \sqrtsnn{0.01} where it depends linearly on collision energy, and the second region at higher energies where particle production is proportional to \Ethird.
We emphasize that the curves for $f \snn$ and $f \Ethird$ are not exact fits and the constant factor $f$ was fiducially chosen and it is the same in the two \snn\ regions.
The function derived by the PHOBOS Collaboration proportional to $\ln{\snn}^2$ \cite{Alver:2010ck} is fitted to data points from the PHENIX Collaboration \cite{Adare:2015bua}.
We note that to properly map the \Ethird\ dependence one would have to study the data in similar bins of \Npart\ instead of the slices in collision centrality. This at present is not possible but $f$ could be extracted from fits to the \Npart\ dependence of multiplicity at each \snn.
While centrality selection is proportional to \Npart\ the evolution of this dependency with \snn\ is not necessarily negligble.
Nevertheless, the \Ethird\ does well to explain the trends seen in data.

\begin{figure}[tb]
\begin{center}
\includegraphics[width=8cm]{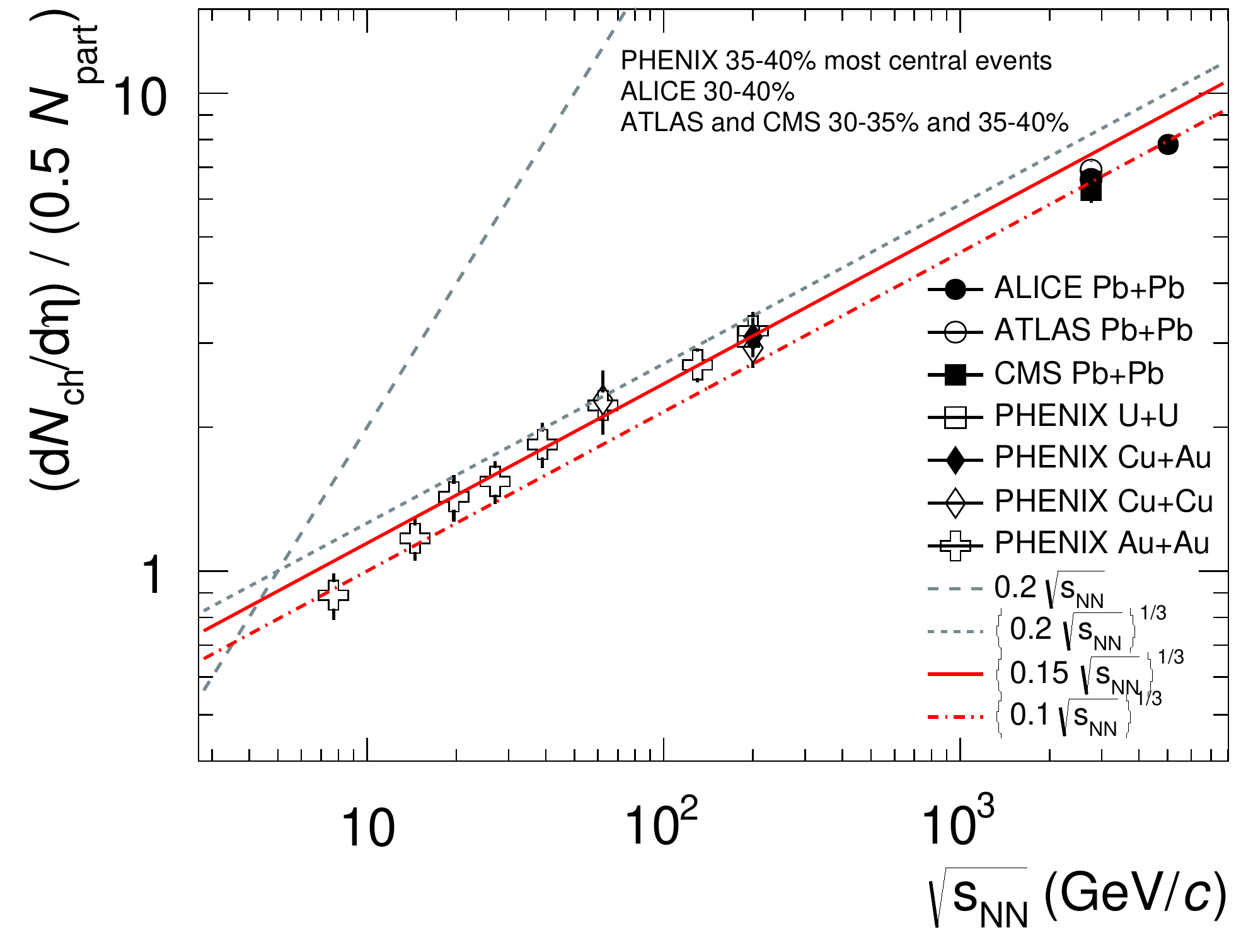}
\caption{Excitation function of pseudorapidity density of charged particles per participating nucleon pair in semi-central heavy-ion collisions \cite{Aamodt:2010cz,Adam:2015ptt,ATLAS:2011ag,Chatrchyan:2011pb,Adare:2015bua}. See text for details.}
\label{logmultperi}
\end{center}
\end{figure}

\begin{figure}[tb]
\begin{center}
\includegraphics[width=8cm]{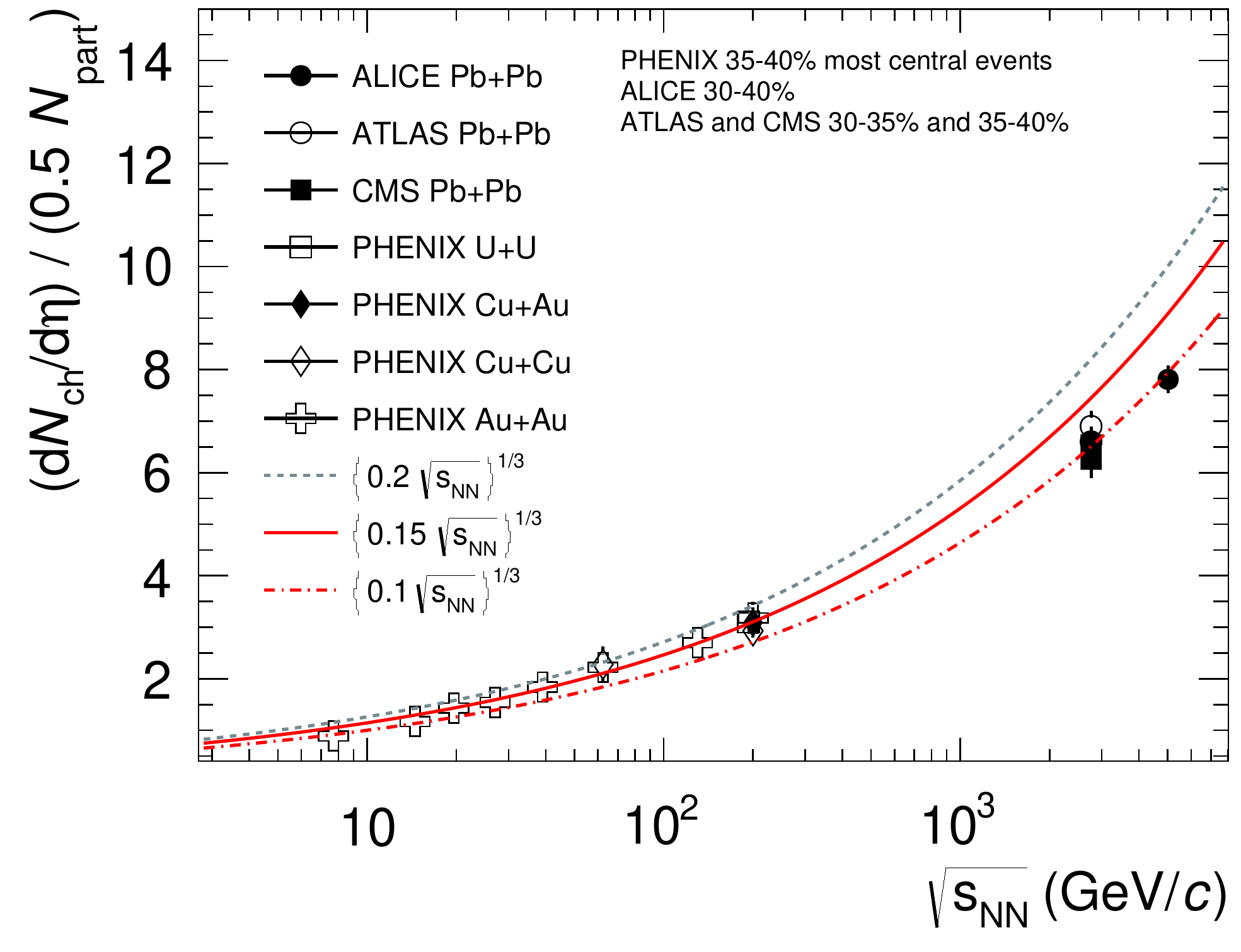}
\caption{Same as Fig. \ref{logmultperi} but shown with linear multiplicity axis emphasizing the comparison of the parametrizations of the excitation functions at the LHC energies.}
\label{linearmultperi}
\end{center}
\end{figure}


In Fig. \ref{logmultperi} and Fig. \ref{linearmultperi} the PHENIX data points correspond to centrality selection of 35-40\%. For CMS and ATLAS two points (for each) are shown: 30-35\% and 35-40\%, while for ALICE the centrality selection is 30-40\%.
The line for $f=0.15$ (with negligible error) is a result of a fit to RHIC data for $\snn > 20$~GeV.
The line for $f=0.1$ is to guide the eye - a good match at the LHC top energy.
Grey dashed lines are to guide the eye and correspond to the lines for most central events shown in Fig. \ref{logmult} and Fig. \ref{linearmult}.

\subsection{On the upper limit of entropy rate}

Taking that entropy production is proportional to particle production, the \Ethird\ dependence of charged particle multiplicity seen in data implies that a limiting rate of entropy production, restricted by hadron formation time, was reached in high-energy heavy-ion collisions.
Moreover, studies performed by PHENIX \cite{Adare:2015bua} and the LHC experiments \cite{Adam:2015ptt} show that the shape of the charged particle multiplicities do not change between \snn\ of 7.7 GeV and 5.02 TeV.
One can consider whether this behavior reflects a general law.
In thermodynamics, it is postulated that system evolves toward maximum entropy.
In the three-dimensional case, maximum possible entropy would translate into particle production proportional to the volume and thus to beam energy.
Instead, it is observed that the maximum achievable particle production and thus entropy grows considerably slower, proportionally to time of expansion into freezeout volume in only one dimension.
The above consideration might not be a definite proof of a new general law, but within the observable physics as described by standard model it appears that what is seen in nucleus-nucleus collisions at RHIC and LHC energies is a maximum achievable rate of entropy production per unit time.
In order to exceed this rate, some new, stronger force would have to exist. That can be never ruled out, even if it is difficult to imagine since it would necessarily have to result in observable phenomena, e.g. at the cosmological scale.
We speculate that the observed scaling of maximum entropy production with time might have consequences also to the physics of early fast expanding universe, e.g. description of inflation stage.

\subsection{On the onset of QGP in hadron collisions}

As noted, based on the above considerations for the transition between \snn\ and \Ethird\ one can draw a general conclusion concerning the formation of QGP in heavy-ion collisions.
However, one should also claim the same for the smaller collision systems.
All LHC experiments observed an intriguing  phenomena of so-called collective particle production in either proton-proton, proton-lead or lead-lead collisions that are characterized by large number of charged particles (high pseudo-rapidity density of produced particles).
Our consideration of \Ethird\ dependence of the particle production indicate that independently of the collision system once the \dNdetaNparthalf\ is about a few or more than 1 after discounting for particle production from hard transverse jets a form of QGP can be present.

Moreover, we project that the search for the end-point at lower energies ($E$ between 4 and 15 GeV) of the second order phase transition can benefit from studies of particle production between the multiplicity bins rather than from tedious variations of the beam energy.
On the other hand, we claim that the most interesting region is at the crossing between the linear dependence and \Ethird\ of the multiplicity density.
At this point judging from the studied cenrality dependence of the \snn\ and \Ethird\ the interesting region is where the $1 \leq \dNdetaNparthalf \leq 1.5$.
From our investigations it lies somewhere between \sqrtsnn{0.007} and \sqrtsnn{0.015}.

Another interesting aspect is a prediction for particle production for the energies much larger than the LHC.
Current considerations of a future accelerators, such as Future Circular Collider \cite{Dainese:2016gch}, consider energies of 40 TeV in the centre-of-mass.
While such an increase in energy does not yet warrant existence of new type of QGP it will be most interesting how strongly the \dNdetaNparthalf\ differs from the \Ethird\ dependence, and how quickly it approaches the dependence that can be extracted from proton-proton collisions.

\subsection{On the excitation function of the transverse energy pseudo-rapidity density}

We note that the transverse energy follows a trend, similar to \Ethird\ scaling \cite{Adare:2015bua}.
It appears that the proportionality factor between $E_{\rm T}$ and $N_{\pi}$ grows relatively slowly ($\sqts^{0.08}$) in the region of 1 GeV.
Such values are comparable to the scale of energy fluctuations corresponding to the radius of strong force and to the pion radius.
However, in the scenario where longitudinal 1D phase-space is filled the transverse energy is not directly related to filling a 1D phase space but might rather be driven by energy fluctuations in the strong field or the initial fluctuations of the transverse energy. Such fluctuations in the transverse direction can be related to the phenomenology of inital stages using strong classical field approach and potential instabilities found by studying the expanding color flux tubes \cite{Fujii:2008dd}, and thus, may be related to the saturation scale of the colliding nuclei.
On the other hand, one can speculate that for the inclusive measurements of multiplicity (including the hard jet fragmenation) while the rate of particle production is fixed the growth of the collision energy (thus relative contribution of hard processes) will cause a steeper growth of the energy density.

\section*{Acknowledgements}
This work is supported by the Slovak Research and Development Agency under contract
APVV-15-0225, and by the U.S. Department of Energy, Office of Science, Office of Nuclear Physics, under contract DE-AC02-05CH11231.

\bibliographystyle{apsrev4-1}
\bibliography{dsdt}
\end{document}